# Spin-dependent Optical Excitations in LiFeO$_2$

Vo Khuong Dien, Nguyen Thi Han,* Wu-Pei Su, and Ming-Fa Lin*

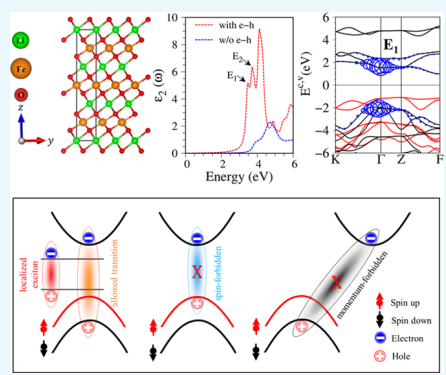

**ABSTRACT:** The three-dimensional ternary LiFeO$_2$ compound presents various unusual properties. The main features are thoroughly explored by using many-body perturbation theory. The concise physical/chemical picture, the critical spin polarizations, and orbital hybridizations in the Li−O and Fe−O bonds are clearly examined through geometric optimization, quasi-particle energy spectra, spin-polarized density of states, spatial charge densities, spin-density distributions, and strong optical responses. The unusual optical transitions cover various frequency-dependent absorption structures, and the most prominent plasmon modes are identified from the dielectric functions, energy loss functions, reflectance spectra, and absorption coefficients. Optical excitations are anisotropic and strongly affected by excitonic effects. The close combinations of electronic, magnetic, and optical properties allow us to identify the significant spin polarizations and orbital hybridizations for each available excitation channel. The lithium ferrite compound can be used for spintronic and photocatalysis applications.

## 1. INTRODUCTION

The lithium ferrite compound (LiFeO$_2$) has been extensively studied for its versatile properties and practical applications. Indeed, LiFeO$_2$ is a non-toxic compound, which exhibits a rapid lithium transportation,[1] a very strong reaction with CO$_2$ molecules,[2,3] magneto-electronic effects,[4−6] and high photocatalysis performances.[7] These features make it attractive in technological applications, such as lithium batteries, catalysis, CO$_2$ absorbers, and spin-electronic technology. Depending on the conditions of preparation, LiFeO$_2$ can survive in different structural forms,[8] that is, as an α-LiFeO$_2$-type cubic rock salt structure,[6] a γ-LiFeO$_2$ rhombohedral ordered rock salt structure,[9,10] and a α-NaFeO$_2$-type rhombohedral structure.[5,11] Among these, the rhombohedral type of LiFeO$_2$ with a layered structure is one of the most common configurations.

Given the vast potential applications of LiFeO$_2$ compounds, our knowledge about their fundamental properties is rather limited. (i) Previously, X-ray photoelectron spectroscopy has been successfully used to investigate the chemical state and the overall valence electronic structure of the LiFeO$_2$ compound.[12] However, we are not aware of any experiments conducted to evaluate the electronic band gap. A few theoretical calculations are based on density functional theory (DFT)[5,13] and neglect the electron−electron self-correlation effects and thus yield inadequate values of the electronic band gaps. For example, the standard DFT calculation of Kalantarian et al. indicated that layered LiFeO$_2$ exhibits a metallic nature.[13] After application of DFT + U to treat the strong on-site Coulomb interactions of the localized electrons, the electronic band gap has expanded, with a typical value of 1.2 eV.[5] (ii) Only a few optical absorption measurements,[7,14] however, focus mainly on the threshold frequency rather than the entire energy of the absorption spectrum. Transmissions, electron energy loss functions, and reflection measurements to thoroughly understand the optical properties of the ferrite compounds are absent. Furthermore, previous theoretical investigations on the optical properties of similar substances (AgFeO$_2$ and CuFeO$_2$)[15] have ignored the electron−hole interactions so that accurate results cannot be obtained. Most first-principles studies also do not propose the optical mechanism, for example, the close connections between the electronic, magnetic, and optical properties. (iii) The magnetic properties of LiFeO$_2$ have been frequently studied experimentally, for example, the magnetization of layered LiFeO$_2$ was measured using a SQUID magnetometer and neutron diffractions.[4,16,17] However, to the best of our knowledge, magnetic first-principles calculations for this compound are rather limited.[5,18]

To better understand the fundamental properties of the lithium ferrite compound, a theoretical framework is applied to comprehend the electronic, magnetic, and optical excitations of the layered LiFeO$_2$. This strategy is based on the state-of-the-art first-principles many-body calculations[19] on an optimized structure with position-dependent chemical bonding, the spin-dependent energy band structure, the spatial spin densities and charge density distributions/charge density differences due to various orbitals, the atom- and orbital-projected density of the state related to spin directions and orbital overlaps, and optical responses including the excitonic effects. The energy-decomposed single-/multi-orbital hybridizations and the spin polarizations are utilized to account for the optical onset frequency, various prominent absorption structures, extraordinary plasmon responses in terms of the dielectric functions, energy loss functions, reflectance spectra, and absorption coefficients under distinct electric polarizations. The current study is of paramount importance not only for basic sciences but also for high-tech applications, for example, spintronic and photo-catalysis applications. Most of the theoretically predicted results in this work can be examined by high-resolution experiments.

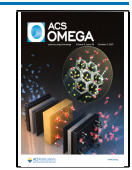

## 2. COMPUTATIONAL DETAILS

**2.1. Ground-State Properties.** In this work, we consider a simple ferromagnetic order in the rhombohedral phase and neglected the antiferromagnetic one observed only at very low Neel temperatures ($T_N$ = 20 K).[20,21] The DFT calculation via the Vienna Ab-initio Simulation Package (VASP)[22] was utilized to estimate the ground-state features of the lithium ferrite compound. To approximate the exchange and correlation potentials, we used the Perdew−Burke−Ernzerhof (PBE) method[23] in the generalized gradient approximation. We employed the projected augmented wave (PAW) pseudopotentials to describe the electronic wave functions in the core region.[24] The cutoff energy for the expansion of the plane wave basis was set to 600 eV. The Brillouin zone was integrated with a special $k$-point mesh of 25 × 25 × 25 in the Monkhorst−Pack sampling technique for geometric optimization. The convergence condition of the ground state was set to $10^{-8}$ eV between two consecutive simulation steps, and all atoms were allowed to fully relax during the geometric optimization until the Hellmann−Feynman force acting on each atom was smaller than 0.01 eV/Å.

**2.2. Quasiparticle Calculations.** Based on the ground-state Kohn−Sham wave functions and the corresponding eigenvalues ($E_{KS}$) of the DFT level, the quasiparticle energy spectrum was obtained by using the GW approximation on the exchange−correlation self-energy. We obtained up to seven self-consistent updates for the quasi-particle Green's function ($G_7W_0$) and found that five iterations ($G_5W_0$) already gave a good convergence of the quasi-particle band gap (See Figure S1). The screening effects are described by using the plasmon-mode model of Hybertsen and Louie.[25] These approaches adopt a cutoff energy of 600 eV for expansion of plane waves and a cutoff energy of 300 eV for the response functions. The Brillouin zone was integrated with a special $k$-point mesh of 15 × 15 × 15 in the Γ sampling technique. The electronic quasi-particle band structure of LiFeO$_2$ was achieved under WANNIER90 codes.[26]

**2.3. Optical Excitations.** Under the perturbation of an electromagnetic wave, the electrons are vertically excited from the occupied states to the unoccupied ones in the quasi-particle energy spectrum. The interactions between photons and the charge carriers of the systems can be well characterized using the macroscopic dielectric functions $\varepsilon(\omega)$. This frequency-dependent function is very useful in understanding the main features of the energy loss functions and reflection and absorption coefficients.

According to Fermi's golden rule,[27] the probability for single-particle excitations can be expressed using the imaginary part of the dielectric function

$$\varepsilon_2^{\uparrow(\downarrow)}(\omega) = \frac{8\pi^2 e^2}{\omega^2} \sum_{vck} |e. \langle v\mathbf{k}^{\uparrow(\downarrow)}|v|c\mathbf{k}^{\uparrow(\downarrow)}\rangle|^2 \times \delta(\omega - E_{c\mathbf{k}}^{\uparrow(\downarrow)} - E_{v\mathbf{k}}^{\uparrow(\downarrow)})$$

where the transition energy is associated with the prominent peak in the optical spectrum and the oscillation strength of each excitation peak is directly related to the joint density of states (DOS) $\delta(\omega - E_{c\mathbf{k}}^{\uparrow(\downarrow)} - E_{v\mathbf{k}}^{\uparrow(\downarrow)})$ and the square of the velocity matrix element, $|e.v\mathbf{k}^{\uparrow(\downarrow)}|v|c\mathbf{k}^{\uparrow(\downarrow)}|^2$.

In general, the electron−hole pairs could be bound through Coulomb interactions under suitable conditions, for example, low screening effects with significant Coulomb electron−electron interactions. The strongly correlated electron−hole pairs, called excitons, are expected to dominate the absorption spectra of the materials. These electron−hole states can be expressed using the expression

$$|S^{\uparrow(\downarrow)}\rangle = \sum_{ks}\sum_v^{hole}\sum_c^{elec} A_{vc\mathbf{k}}^{s\uparrow(\downarrow)}|vc\mathbf{k}^{\uparrow(\downarrow)}\rangle$$

The amplitude $A_{vc\mathbf{k}}^{s\uparrow(\downarrow)}$ is determined by solving the standard Bethe−Salpeter equation (BSE)[28]

$$(E_{c\mathbf{k}}^{QP\uparrow(\downarrow)} - E_{v\mathbf{k}}^{QP\uparrow(\downarrow)})A_{vc\mathbf{k}}^{s\uparrow(\downarrow)} + \sum_{v'c'\mathbf{k}'}\langle vc\mathbf{k}^{\uparrow(\downarrow)}|K^{eh}|v'c'\mathbf{k}'^{\uparrow(\downarrow)}\rangle A_{v'c'\mathbf{k}'}^{s\uparrow(\downarrow)} = \Omega^{s\uparrow(\downarrow)}A_{vc\mathbf{k}}^{s\uparrow(\downarrow)}$$

where $E_{c\mathbf{k}}^{QP\uparrow(\downarrow)}$ and $E_{v\mathbf{k}}^{QP\uparrow(\downarrow)}$ are the excitation energies for the conduction band states and the valence band states, respectively, $K^{eh}$ is the kernel describing the correlated electron−hole pairs, and $\Omega^s$ is the energy of the exited state.

In this step, we used the same $k$-point sampling, cutoff energy, and number of bands as in the GW calculations. Moreover, 11 highest occupied valence bands (VBs) and 6 lowest unoccupied conduction bands (CBs) are included as a basis for the excitonic states with a photon energy region from 0 to 25 eV. The broadening parameter was set equal to 0.1 eV for all optical spectra.

The presence of electron−hole pairs may significantly affect the main features of the optical spectra. The close connection between the quasi-particle charges, spin-polarizations, and orbital hybridizations and the effects of the coupled quasi-particles on optical excitations will be discussed in detail.

## 3. RESULTS AND DISCUSSIONS

**3.1. Electronic and Magnetic Properties.** The LiFeO$_2$ compound crystallizes in a trigonal structure with the $\overline{R}3m$ space group (Figure 1a. The calculated lattice constants of this structure (Table 1) are 2.88, 2 .88, and 14.31 Å in $x$, $y$, and $z$ directions, respectively, which are in very good agreement with previous theoretical and experimental values.[11,29−31] The basic building block of LiFeO$_2$ consists of alternately stacked FeO$_2$ and Li layers along the $z$-axis. Each Fe/Li atom occupies the center of an octahedron of O atoms (Figure 1b). The highly ordered arrangement of the atoms and the anisotropy of the geometric structure, which originate from the complex orbital hybridizations, are responsible for the anisotropic behavior of the optical excitations.

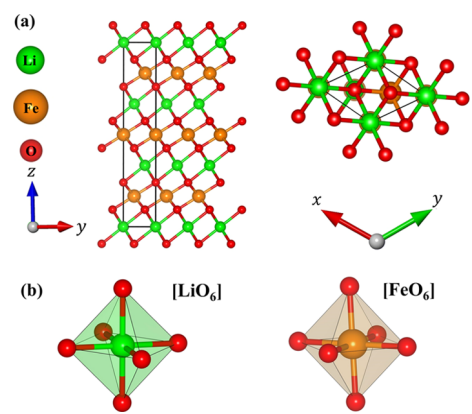

Figure 1. Optimal geometric structure of the LiFeO$_2$ compound. (a) Side view and top view and (b) octahedron structure of [LiO$_6$] and [FeO$_6$].

Table 1. Fundamental Band Gap and Magnetic Moment Computed Using Various Approaches

| approach | electronic band gap (eV) | magnetic moment ($\mu_B$) | | | |
|---|---|---|---|---|---|
| | | Li | Fe | O | tot |
| DFT[a] | Metallic | 0.01 | 3.71 | 0.43 | 4.15 |
| DFT[b] | Metallic | | | | |
| DFT + U[a] | 1.22 | 0.01 | 4.22 | 0.48 | 4.71 |
| DFT + U[b] | 1.2 | | | | 4.95 |
| $G_7W_0$[a] | 1.9 | 0.01 | 4.06 | 0.55 | 4.62 |

[a]This work. [b]Reference 5 13,.

The energy band structure along the high-symmetry point of the LiFeO$_2$ compound was calculated and is shown in Figure 2a, in which the Fermi level is set at the middle of the CB and VB. Many energy sub-bands with various dispersion characteristics, such as oscillatory, camel's back shape, or dispersionless relations, owing to the contribution of certain orbitals and atoms in the unit cell, are present. The spin-dependent energy band structure of the LiFeO$_2$ compound is expressed explicitly, especially the remarkable spin splitting in the vicinity of the Fermi energy. For the spin-up and spin-down states, the energy bands nearest to the $E_F$ are fully occupied and unoccupied, respectively, which leads to a direct gap of 1.9 eV at the F symmetry point. Very interestingly, the energy band gaps for spin-up and spin-down states are extremely different; typical values are about 2.8 and 5.8 eV for the former and the latter, respectively. The large spin splitting with a complicated energy band structure reflects the ferromagnetic configuration of the LiFeO$_2$ compound and is responsible for the unusual optical properties.

The orbital hybridizations and spin polarizations in the Li–O and Fe–O bonds could be clarified using the atom and orbital DOS. As revealed in Figure 2b, the DOS of the 3D ternary LiFeO$_2$ compound is characterized by symmetric sharp peaks, the shoulder structures, and a plateau peak, resembling that of the band edge in 1D because they mainly originated from parabolic, the camel's back shape, and linear energy sub-bands, respectively. The merged DOS of difference atoms and orbitals provides the complicated orbital hybridizations in Li–O and Fe–O bonds. Interestingly, the domination and energy of the spin-up and spin-down states are very different. This indicates that LiFeO$_2$ possesses a very strong ferromagnetic behavior, which is in good agreement with the spin-splitting band structure and the spin-density distribution (discussed later). According to the distribution of atoms and orbitals (Figure 2b), the electronic structures of the LiFeO$_2$ compound can be classified into four categories: (i) 6.5 eV ≤ $E^{c,v}$ ≤ 14.7 eV is due to Li 2s and O 2p orbital and Fe 4s and O 2p orbital hybridizations, (ii) 1 eV ≤ $E^{c,v}$ ≤ 5 eV is a mixture of Li 2s and O 2p orbital and Fe 3d and O 2p orbital hybridizations, (iii) −9 eV ≤ $E^{c,v}$ ≤ −1 eV is related to the coupling of Li 2s and O 2p orbitals and Fe 3d and O 2p orbitals, and (iv) $E^{c,v}$ ≤ −21 eV is strongly dominated by O 2s and a minority of Li 2s and Fe 3d orbitals. The significant hybridization of O 2p and Fe 3d compositions in the occupied state and the strong domination of O 2s below −21 eV are consistent with the X-ray photoelectron measurements[12] (Figure S3).

The quasiparticle corrections to the Kohn–Sham eigenvalues are plotted in Figure 2c. After taking the e–e self-energy effects into account, the valence and conduction states are strongly modified. Especially, for the Fe 3d↓ states near the Fermi level (the dashed blue oval), the significant rising up of these states leads to metal−semiconductor transitions. Compared with DFT[13] and DFT + U[5] (Figure S2), the wider energy gap suggests an enhancement of Coulomb interactions and the low screening makes the excitonic effects explicit. Furthermore, the Coulomb screening effects of the spin-up and spin-down states are also significantly different. Apparently, due to the rather complicated quasiparticle corrections, the quasiparticle band structure may not be achieved using a simple "scissor" operator.

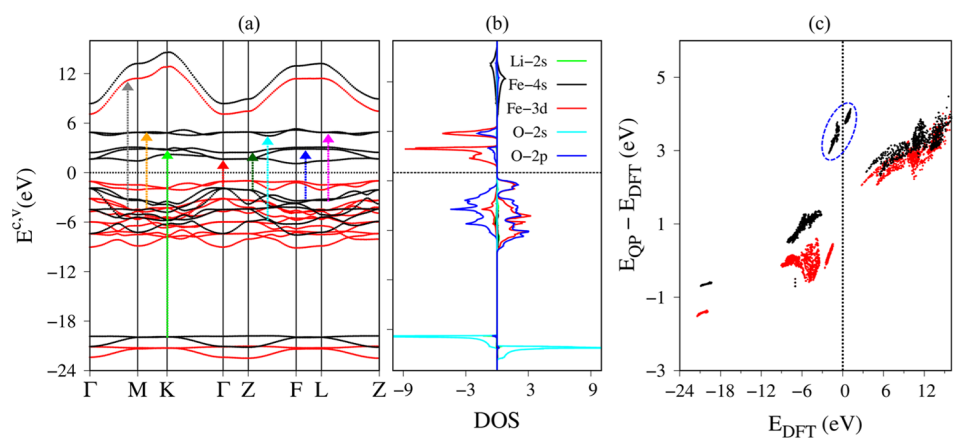

Figure 2. (a) $G_7W_0$ quasi-particle band structure along the high-symmetry points in the wave-vector space; the red/black lines represent the spin-up/spin-down states and the vertical colored arrows indicate the optical excitations. (b) Orbital-projected density of states and (c) differences between the $G_7W_0$ quasi-particle energy and the PBE Kohn−Sham eigenvalue of the LiFeO$_2$ compound; the red/black dots represent the spin-up/spin-down states and the Fermi level is denoted by the dashed black line.

Bader charge analysis and the spatial charge density distribution/charge density variation (Figure 3a–d) can further provide information about the character of the Li–O and Fe–O chemical bonds. Bader charge analysis shows that Li and Fe atoms transfer their charges to the O atoms because the average effective charges are determined to be 0.98 e for a Li atom, 1.08 e for a Fe atom, and −1.03 e for an O atom. As for the charge distribution, the inner and the outer regions of the O atom arise from 2s and 2p orbitals, respectively. Similarly, the inner and outer parts of the Fe atom correspond to the 4s and 3d orbitals. The serious distortion of the spherical charges and significant charge variations clearly suggest an ionic–covalent nature of the Fe–O bond due to significant Fe (4s, 3d)–O (2p) orbital hybridizations. As for the O 2s orbital, it almost does not reacts at all ($\Delta\rho \approx 0$ at the innermost of O atom) because its ionization energy is the largest among all orbitals (lowest energy and almost do not hybridize with other orbitals in DOS). Considering the Li–O chemical bond, the charge density around the Li atom is contributed by the 2s orbital. The slight deformation of the Li outer sphere and its valence charges are almost transferred to the O atom, illustrating the ionic nature of the hybridization of Li 2s and O (2s and 2p) orbitals.

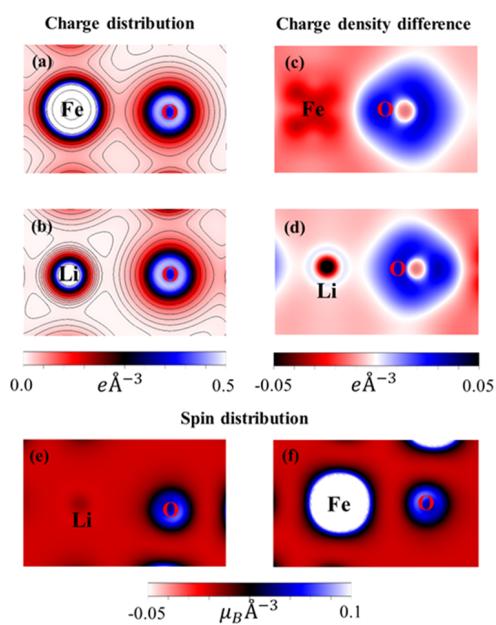

**Figure 3.** (a,b) charge density distribution, (c,d) charge density difference, and (e,f) spin-density distribution related to the significant orbital hybridizations in the Fe–O and Li–O bonds, respectively.

The magnetic properties of the lithium ferrite compound could be comprehended through the spin density distribution (Figure 3e,f) and the magnetic moment (Table 1), in which the net magnetic moment in a unit cell is determined by the competition between spin-up and spin-down components. The LiFeO$_2$ compound is ferromagnetically ordered, and the total magnetic moment is 4.62 $\mu_B$, as expected, because the Fe atom introduces four spin-up electrons. As a matter of fact, the spin-up density is the most dominant part and relies on the Fe atom (4.06 $\mu_B$), which is directly reflected by the fully occupied states in the strongly dispersive energy bands below the Fermi level (the third region). Furthermore, because of the strong hybridization with the Fe atom, O and Li atoms also manifest partial minor magnetic contributions with values one and two orders lower than those of the Fe atom, respectively. As a result, the net magnetic moments are expected to be sensitive to change during Li$^+$ transportation.[5] The ferromagnetic configuration, the magnetic moment of the Fe$^{3+}$ ion, and the net magnetic moment of LiFeO$_2$ have been confirmed in previous neutron diffraction[4] and first-principles calculations.[5,31]

**3.2. Optical Properties.** The dielectric function expresses the main electronic and magnetic properties. The optical excitations of the lithium ferrite compound are described using the imaginary part $[\varepsilon_2(\omega)]$ of the dielectric function. As presented in Figure 5, the optical properties of lithium ferrite express twofold degeneracies; the calculated $\varepsilon_2(\omega)$ for the $x$ and $y$ polarizations are the same, but the curve for the $z$ polarization is totally different, owing to the different atomic arrangement. This indicates that the trigonal LiFeO$_2$ exhibits a strong optical anisotropy.

In the absence of electron–hole couplings, the optical gap is situated at 3.9 eV for $z$-polarizations (Figure 6a). The rapid oscillations originate from the direct transition at the Γ point of the spin-down states between O 2p and Fe 3d orbitals (the red arrow in Figure 2a). Notice that the indirect/difference spin transitions are forbidden due to the lack of required momentum transfer/spin flip under photon absorption (Figure 4). For example, the promotion of electrons from the highest occupied to the lowest unoccupied states at the F point is forbidden due to the opposite spin orientations. Beyond the threshold frequency, $\varepsilon_2(\omega)$ also possesses many special structures with a variety of shapes. For instance, the singularity 4.7 eV with a relatively strong response is related to the first valence and the third conduction sub-bands at the Z point due to the multi-orbital of O 2p → Fe 3d and O 2p → O 2p of the spin-down state (the green arrow in Figure 2a). The transition along the FL path by the multi-orbital of O 2p → Fe 3d and O 2p → O 2p of the similar spin state leads to the prominent 7.0 eV peak (the blue arrow in Figure 2a); the observable 8.3 eV excitation structure is associated with the transitions from the occupied to unoccupied ones along the LZ path through O 2p → Fe 3d and O 2p → O 2p of the spin-down state (the pink arrow in Figure 2a); the absorption structure at 9.2 eV is related to the sub-bands along the MK path by the multi-orbital of O 2p → Fe-3d and O-2p → O-2p of the spin-down state (the orange arrow in Figure 2a); a weak but significant 11.1 eV prominent peak is related to the multi-orbital of O 2p → Fe 3d and O 2p → O 2p due to the transition of the spin-down state along the ZF path (the cyan arrow in Figure 2a); the absorption peak at 23.1 eV is due to the transition of the spin-down state under the orbitals of O 2s → Fe 3d (the green arrow in Figure 2a). Very interestingly, the promotion of electrons with spin-up states only creates one prominent peak at 16 eV (the gray arrow in Figure 2a) because of only one prominent Fe 4s van Hove singularity in the conduction bands (few available transitions), and the oscillation strengths are rather weak for the higher-energy transitions. The macroscopic orbital characters and spins of each channel have been assigned and shown in Table 2.

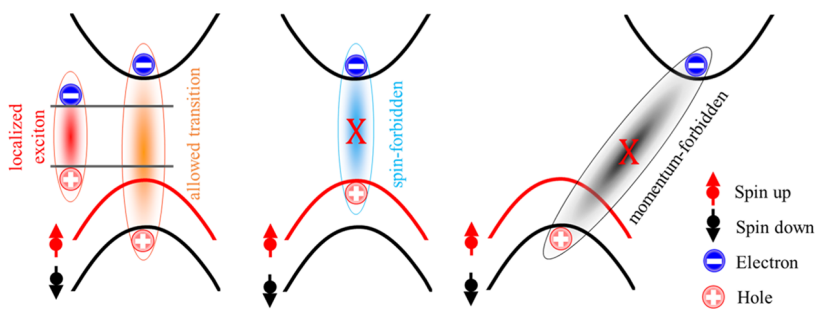

**Figure 4.** Optical excitation mechanisms of the LiFeO$_2$ compound. The Coulomb mutual interactions of excited electrons–holes are called excitons. Allowed transitions consist of electrons and holes located at the same wave vector in the reciprocal space with the same spin orientation (bright excitons). Momentum-forbidden transitions involve electrons and holes located at different band edge states in the reciprocal space. Spin-forbidden transitions involve electrons and holes with opposite spins. The momentum- and spin-forbidden transitions cannot be excited by electromagnetic wave (dark excitons) due to the lack of a required momentum transfer and spin flip, respectively.

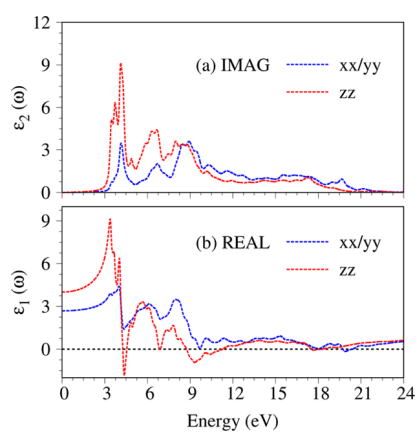

**Figure 5.** (a) Imaginary and (b) real parts of dielectric functions with the excitonic effects under three electronic polarizations.

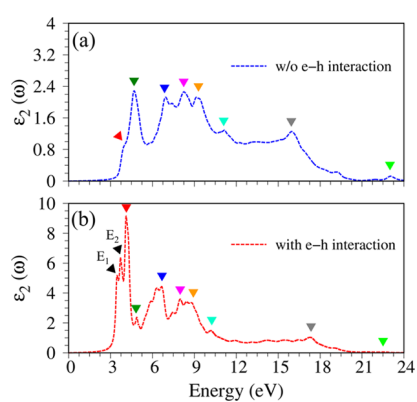

**Figure 6.** Comparison of the imaginary parts of the dielectric function in the z-direction of (a) with and (b) without excitonic effects. The colored triangles assign the prominent optical excitations, corresponding to the colored arrows in Figure 2a.

Specifically, the optical excitations, which include the excitonic effects, are also shown in Figure 6b. The optical gap is reduced with a typical exciton binding energy of 0.6 eV, and the excitation strength/energy is enhanced/changed as a consequence of electron−hole wave function overlaps (Figure 6b and Table 2).[32] The red shift of the optical gap and the remarkable change in these curves indicate a significant impact of the excitonic effects on the optical properties of the $LiFeO_2$ compound. The nature of specific exciton peaks is also understood from the distribution of the electron−hole pair wave functions.[33,34] For example, the first two peaks located at 3.3 eV ($E_1$) and 3.8 eV ($E_2$) (Figure 7a) are clearly excitonic in nature because the optical absorption without electron−hole couplings nearly vanishes in the interval of 0 to 4 eV. These sharp peaks are made of the electron−hole coherent superposition, emanating from the vertical transitions between the last occupied state and the first-two unoccupied states of the spin-down band structure. The contributions to the former and the latter excitons arise mostly from the band extrema vicinity of the Γ and Z points, respectively, where the dispersion of the valence sub-bands is almost parallel to the conduction one (Figure 7b,c). Obviously, owing to the difference in band structures, the electrons with different spin orientations have dramatically different dielectric

**Table 2. Calculated Prominent Absorption Structures and the Leading Transition of Each Peak**

| | energy (eV) | | | |
|---|---|---|---|---|
| colour | w/o e−h | with e−h | spin | orbital hybridizations leading to excitations (Fe−O) |
| black | | 3.3 | ↓ | O 2p → Fe 3d |
| black | | 3.8 | ↓ | O 2p → Fe 3d |
| Red | 3.9 | 3.92 | ↓ | O 2p → Fe 3d |
| green | 4.7 | 4.73 | ↓ | O 2p → Fe 3d |
| blue | 7.0 | 6.7 | ↓ | O 2p → Fe 3d & O 2p → O 2p |
| pink | 8.3 | 8.0 | ↓ | O 2p → Fe 3d & O 2p → O 2p |
| orange | 9.2 | 8.9 | ↓ | O 2p → Fe 3d & O 2p → O 2p |
| cyan | 11.1 | 10.7 | ↓ | O 2p → Fe 3d & O 2p → O 2p |
| gray | 16.0 | 17.5 | ↑ | O 2p → Fe 4s & Fe 3d → Fe 4s |
| light-green | 23.1 | 22.6 | ↓ | O 2s → Fe 3d |

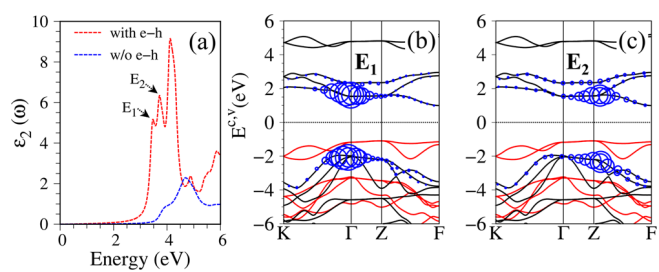

**Figure 7.** (a) Enlarged imaginary part of the dielectric function in the $z$-direction of with and without excitonic effects. (b) Amplitude of the first ($E_1$) and (c) second ($E_2$) exciton peaks is plotted in fat-band styles. The radii of the circles represent the contribution of electron–hole pairs at the $k$-point to the ith exciton wave function. The solid red/black lines in the background are the corresponding G7W0 quasi-particle band structures of spin-up/down states, and the dashed line represents the Fermi level.

responses. The unique optical excitations of $LiFeO_2$ can be exploited for spintronic applications.[14]

In addition to the single-particle excitations, the collective excitations are also expected to dominate the coupling of charges and electromagnetic waves. The energy loss function, defined as $\text{Im}[-1/\varepsilon(\omega)]$ (Figure 8a), is useful to elucidate the coherent oscillation of the valence charges, in which, each prominent peak is referred to as a plasmon, the collective excitations of free charges appear at frequencies where $\varepsilon_1(\omega)$ passes through zero and $\varepsilon_2(\omega)$ is nearly vanishing (Figure 5). For the opposite cases, the finite $\varepsilon_2(\omega)$ implies that the plasmon modes might decay into single particle excitations, owing to the Landau damping. There are two intrinsic peaks at 17 and 18 eV in the screened excitation spectrum in the $x/y$ and $z$ directions, respectively. These are mainly attributed to the excitation of Li 2s, O 2p, and Fe (4s and 3d) orbitals; the contribution of the 2s orbital of the O atom to these peaks is ignored because it contributes only minimally in the active regions. Furthermore, the energy loss function also presents several relatively weak plasmon modes because the zero points disappeared or are accompanied by extreme Landau dampings.

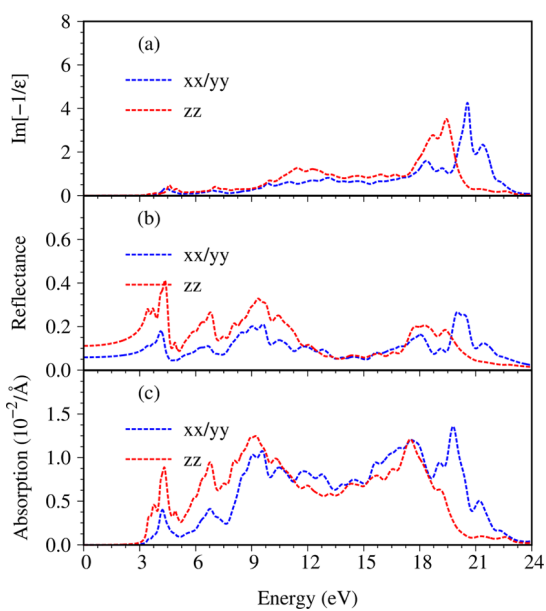

**Figure 8.** Various optical properties: (a) energy loss functions, (b) reflectance spectra, and (c) absorption coefficients under $x/y$ and $z$ electric-polarizations.

When a beam of photons arrive at the boundary of the material, some of them will be reflected from the front surface, while others will transmit and incorporate with the valence electrons through many interband transitions. The inverse of the absorption coefficient $[\alpha(\omega)]$ is the characteristic length through which an electromagnetic wave can penetrate a condensed-matter system. The various kinds of valence/conduction electrons are able to efficiently screen the electromagnetic wave at different frequency ranges, or such carriers can generate diversified optical excitations. The significant couplings with an electromagnetic wave absorption directly determine the reflectance and absorption spectra. Considering the low-energy region, the reflection is weakly dependent on the energy, with typical values of 10 and 6% in the $x/y$ and $z$ directions, respectively (Figure 8b). The absorption coefficient vanishes because of the absence of optical excitations (Figure 8c). However, due to the excitation events, the reflection could change sensitively and exhibits a large fluctuation, especially in the $z$-direction (∼30%) in the strongest plasmon mode, owing to the influence of the coherent excitations. Similarly, for the absorption coefficients, the absorption at 4 eV is strong, as expected for spin-allowed transitions. The presence of large variation is a consequence of the optical excitations due to various orbital and spin contributions. The inverse value of the absorption coefficient mostly lie in a range of 67–200 Å, meaning that photon beams penetrating the $LiFeO_2$ medium will be easily absorbed through the rich optical excitations, and thus, the lithium ferrite compound has potential applications in photo-catalysis technology. Far away from the plasmon energies, the photons illuminating $LiFeO_2$ medium are almost transmitted because the electrons are unable to respond fast enough to screen such incident light.

The results of the photoluminescence, adsorption, reflection, or transmission measurements,[35−38] being supported by the Kramers−Kronig relations,[39] can be compared with our dielectric function, absorption, reflection coefficients, and energy loss functions. Other information, such as the diversified optical excitation peaks, the redshift of the optical gap due to an excitonic effect, and the strong anisotropy of the optical responses can also be examined. Apparently, our analysis of spin polarizations and orbital hybridizations related to optical properties is very difficult to achieve in experiments.

## 4. CONCLUSIONS

In the current work, first-principles calculations with many-body perturbation theory were used to investigate the geometric, electronic, magnetic, and optical properties of $LiFeO_2$. As a result, the relation between these properties and the spin polarizations and orbital hybridizations can be successfully understood. The state-of-the-art analysis is very useful for fully comprehending the diversified properties in magnetic systems and other emerging materials.

The 3D $LiFeO_2$ compound presents very unique features, for example, a unit cell with highly organized atoms and ions, anisotropic geometry, spin-polarized band structure, spatial spin and charge redistribution, and many van Hove singularities due to the spin polarizations and extreme point dispersions. The band edge states with orbital character and spin orientations that are assigned to the optical excitations can be well characterized. Regarding the optical properties, the robust and important impact of excitonic effects are revealed through a significantly reduced optical gap and a large enhancement of the adsorption intensity. The optical responses involve low-reflectance/high-transmission coefficients at energies lower than the optical gap, the presence of various prominent excitation peaks, a large prominent plasmon mode with energy beyond 17 eV due to the contribution of certain valence electrons, and the sensitivity of the absorption and reflectance spectra on the excitation events. The developed theoretical framework could be extended to other ferrite compounds.

## ■ ASSOCIATED CONTENT

### ⓈSupporting Information

The Supporting Information is available free of charge at https://pubs.acs.org/doi/10.1021/acsomega.1c03698.

Convergence of the GW gap, band structure comparison (DFT, DFT + U, and G7W0), DOS comparison (DFT, DFT + U, G7W0, and experimental), and computational details for optical properties ([PDF](#))

## ■ AUTHOR INFORMATION

### Corresponding Authors

Nguyen Thi Han − *Department of Physics, National Cheng Kung University, Tainan 701, Taiwan*;
Email: han.nguyen.dhsptn@gmail.com

Ming-Fa Lin − *Department of Physics, National Cheng Kung University, Tainan 701, Taiwan; Hierarchical Green Energy Materials, Hi-research Center, National Cheng Kung University, Tainan 701, Taiwan;* ⓘ orcid.org/0000-0002-0531-2068; Email: mflin@mail.ncku.edu.tw

### Authors

Vo Khuong Dien − *Department of Physics, National Cheng Kung University, Tainan 701, Taiwan;* ⓘ orcid.org/0000-0002-7974-9852

Wu-Pei Su − *Department of Physics and Texas Center for Superconductivity, University of Houston, Houston, Texas 77204, United States*

Complete contact information is available at:
https://pubs.acs.org/10.1021/acsomega.1c03698

### Author Contributions

V.K D. designed the study and wrote the manuscript. N.T.H. performed the simulations and analyzed the results. W.P.S discussed the data and revised the manuscript. M.F.L. supervised the work. All authors discussed the results and revised the manuscript.

### Notes

The authors declare no competing financial interest.

## ■ ACKNOWLEDGMENTS

This work was financially supported by the Hierarchical Green-Energy Materials (Hi-GEM) Research Center, the Featured Areas Research Center Program within the framework of the Higher Education Sprout Project by the Ministry of Education (MOE), and the Ministry of Science and Technology (MOST 110-2634-F-006 -017) in Taiwan.

## ■ REFERENCES

(1) Armstrong, A. R.; Tee, D. W.; La Mantia, F.; Novák, P.; Bruce, P. G. Synthesis of Tetrahedral LiFeO2 and Its Behavior as a Cathode in Rechargeable Lithium Batteries. *J. Appl. Comput. Sci.* **2008**, *130*, 3554−3559.

(2) Yanase, I.; Kameyama, A.; Kobayashi, H. $CO_2$ absorption and structural phase transition of α-LiFeO2. *J. Ceram. Soc. Jpn.* **2010**, *118*, 48−51.

(3) Gómez-García, J. F.; Mendoza-Nieto, J. A.; Yañez-Aulestia, A.; Plascencia-Hernández, F.; Pfeiffer, H. New evidences in CO oxidation and selective chemisorption of carbon oxides on different alkaline ferrite crystal phases (NaFeO2 and LiFeO2). *Fuel Process. Technol.* **2020**, *204*, 106404.

(4) Chappel, E.; Holzapfel, M.; Douakha, N.; Chouteau, G.; Ott, A.; Ouladdiaf, B. Magnetic structure of layered LiFe1−xCoxO2: a powder-neutron diffraction study. *J. Magn. Magn. Mater.* **2002**, *242−245*, 738−740.

(5) Boufelfel, A. Electronic structure and magnetism in the layered LiFeO2: DFT+U calculations. *J. Magn. Magn. Mater.* **2013**, *343*, 92−98.

(6) Layek, S.; Greenberg, E.; Xu, W.; Rozenberg, G. K.; Pasternak, M. P.; Itié, J.-P.; Merkel, D. G. Pressure-induced spin crossover in disordered α−LiFeO2. *Phys. Rev. B: Condens. Matter Mater. Phys.* **2016**, *94*, 125129.

(7) Gu, W.; Guo, Y.; Li, Q.; Tian, Y.; Chu, K. Lithium Iron Oxide (LiFeO2) for Electroreduction of Dinitrogen to Ammonia. *ACS Appl. Mater. Interfaces* **2020**, *12*, 37258−37264.

(8) Li, J.; Li, J.; Luo, J.; Wang, L.; He, X. Recent advances in the LiFeO2-based materials for Li-ion batteries. *Int. J. Electrochem. Sci.* **2011**, *6*, 1550−1561.

(9) Barré, M.; Catti, M. Neutron diffraction study of the β′ and γ phases of LiFeO2. *J. Solid State Chem.* **2009**, *182*, 2549−2554.

(10) Barriga, C.; Morales, J.; Tirado, J. L. Structural modifications induced by proton exchange in γ-LiFeO2. *Mater. Res. Bull.* **1990**, *25*, 997−1002.

(11) Shirane, T.; Kanno, R.; Kawamoto, Y.; Takeda, Y.; Takano, M.; Kamiyama, T.; Izumi, F. Structure and physical properties of lithium iron oxide, LiFeO2, synthesized by ionic exchange reaction. *Solid State Ionics* **1995**, *79*, 227−233.

(12) Galakhov, V. R.; Kurmaev, E. Z.; Uhlenbrock, S.; Neumann, M.; Kellerman, D. G.; Gorshkov, V. S. Electronic structure of LiNiO2, LiFeO2 and LiCrO2: X-ray photoelectron and X-ray emission study. *SSCom* **1995**, *95*, 347−351.

(13) Momeni, M.; Yousefi Mashhour, H.; Kalantarian, M. M. New approaches to consider electrical properties, band gaps and rate capability of same-structured cathode materials using density of states diagrams: Layered oxides as a case study. *J. Alloys Compd.* **2019**, *787*, 738−743.

(14) Tangra, A. K.; Lotey, G. S. Synthesis and investigation of structural, optical, magnetic, and biocompatibility properties of nanoferrites AFeO2. *Curr. Appl. Phys.* **2021**, *27*, 103−116.

(15) Ong, K. P.; Bai, K.; Blaha, P.; Wu, P. Electronic Structure and Optical Properties of AFeO (A = Ag, Cu) within GGA Calculations. *Chem. Mater.* **2007**, *19*, 634−640.

(16) Akiyama, R.; Ikedo, Y.; Månsson, M.; Goko, T.; Sugiyama, J.; Andreica, D.; Amato, A.; Matan, K.; Sato, T. J. Short-range spin correlations in β″-LiFeO2 from bulk magnetization, neutron diffraction, and μSR experiments. *Phys. Rev. B: Condens. Matter Mater. Phys.* **2010**, *81*, 024404.

(17) Viret, M.; Rubi, D.; Colson, D.; Lebeugle, D.; Forget, A.; Bonville, P.; Dhalenne, G.; Saint-Martin, R.; André, G.; Ott, F. β-NaFeO2, a new room-temperature multiferroic material. *Mater. Res. Bull.* **2012**, *47*, 2294−2298.

(18) Meyer, A.; Catti, M.; Dovesi, R. Chemical and magnetic ordering derived from ab initio simulations: the case of β′-LiFeO2. *J. Phys.: Condens. Matter* **2010**, *22*, 146008.

(19) Rohlfing, M.; Louie, S. G. Electron-Hole Excitations in Semiconductors and Insulators. *Phys. Rev. Lett.* **1998**, *81*, 2312−2315.

(20) Douakha, N.; Holzapfel, M.; Chappel, E.; Chouteau, G.; Croguennec, L.; Ott, A.; Ouladdiaf, B. Nuclear and Magnetic Structure of Layered LiFe1−xCoxO2 (0≤x≤1) Determined by High-Resolution Neutron Diffraction. *J. Solid State Chem.* **2002**, *163*, 406−411.

(21) Chappel, E.; Holzapfel, M.; Chouteau, G.; Ott, A. Effect of Cobalt on the Magnetic Properties of the LiFe1−xCoxO2 Layered System (0≤x≤1). *J. Solid State Chem.* **2000**, *154*, 451−459.

(22) Hafner, J. Ab-initio simulations of materials using VASP: Density-functional theory and beyond. *J. Comput. Chem.* **2008**, *29*, 2044−2078.

(23) Peng, H.; Perdew, J. P. Rehabilitation of the Perdew-Burke-Ernzerhof generalized gradient approximation for layered materials. *Phys. Rev. B: Condens. Matter Mater. Phys.* **2017**, *95*, 081105.

(24) Blöchl, P. E. Projector augmented-wave method. *Phys. Rev. B: Condens. Matter Mater. Phys.* **1994**, *50*, 17953.

(25) Hybertsen, M. S.; Louie, S. G. Electron correlation in semiconductors and insulators: Band gaps and quasiparticle energies. *Phys. Rev. B: Condens. Matter Mater. Phys.* **1986**, *34*, 5390−5413.

(26) Pizzi, G.; Vitale, V.; Arita, R.; Blügel, S.; Freimuth, F.; Géranton, G.; Gibertini, M.; Gresch, D.; Johnson, C.; Koretsune, T.; Ibañez-Azpiroz, J.; Lee, H.; Lihm, J.-M.; Marchand, D.; Marrazzo, A.; Mokrousov, Y.; Mustafa, J. I.; Nohara, Y.; Nomura, Y.; Paulatto, L.; Poncé, S.; Ponweiser, T.; Qiao, J.; Thöle, F.; Tsirkin, S. S.; Wierzbowska, M.; Marzari, N.; Vanderbilt, D.; Souza, I.; Mostofi, A. A.; Yates, J. R. Wannier90 as a community code: new features and applications. *J. Phys. Condens. Matter* **2020**, *32*, 165902.

(27) Mahan, G. D. *Many-particle physics*; Springer Science & Business Media, 2013.

(28) Nakanishi, N. A general survey of the theory of the Bethe-Salpeter equation. *Prog. Theor. Phys. Suppl.* **1969**, *43*, 1−81.

(29) Boufelfel, A. Electronic structure and magnetism in the layered LiFeO2: DFT + U calculations. *J. Magn. Magn. Mater.* **2013**, *343*, 92−98.

(30) Kanno, R.; Shirane, T.; Kawamoto, Y.; Takeda, Y.; Takano, M.; Ohashi, M.; Yamaguchi, Y. Synthesis, structure, and electrochemical properties of a new lithium iron oxide, LiFeO2, with a corrugated layer structure. *J. Electrochem. Soc.* **1996**, *143*, 2435.

(31) Carlier, D.; Ménétrier, M.; Grey, C. P.; Delmas, C.; Ceder, G. Understanding the NMR shifts in paramagnetic transition metal oxides using density functional theory calculations. *Phys. Rev. B: Condens. Matter Mater. Phys.* **2003**, *67*, 174103.

(32) Chang, E. K.; Rohlfing, M.; Louie, S. G. Excitons and Optical Properties of α-Quartz. *Phys. Rev. Lett.* **2000**, *85*, 2613−2616.

(33) Bokdam, M.; Sander, T.; Stroppa, A.; Picozzi, S.; Sarma, D.; Franchini, C.; Kresse, G. Role of polar phonons in the photo excited state of metal halide perovskites. *Sci. Rep.* **2016**, *6*, 28618.

(34) Zhang, L.; Mao, X.; Matta, S. K.; Gu, Y.; Du, A. Two-Dimensional CuTe2X (X = Cl, Br, and I): Potential Photocatalysts for Water Splitting under the Visible/Infrared Light. *J. Phys. Chem. C* **2019**, *123*, 25543−25548.

(35) Perkowitz, S. *Optical Characterization of Semiconductors: Infrared, Raman, and Photoluminescence Spectroscopy*; Academic Press Inc.: San Diego, CA, **1993**.

(36) Mirabella, F. M. *Internal reflection spectroscopy: theory and applications*; CRC Press, 1992; Vol. 15.

(37) Brydson, R.; *Electron energy loss spectroscopy*; Garland Science, 2020.

(38) Ibach, H.; Mills, D. L. *Electron energy loss spectroscopy and surface vibrations*; Academic press, 2013.

(39) Taft, E. A.; Philipp, H. R. Optical properties of graphite. *Phys. Rev.* **1965**, *138*, A197.